\documentclass[pre,twocolumn,showpacs,floatfix,longbibliography]{revtex4-1}%

\usepackage[english]{babel}
\usepackage[utf8]{inputenc}
\usepackage{amssymb}
\usepackage{amsmath}
\usepackage{graphicx}
\usepackage[colorinlistoftodos]{todonotes}
\usepackage{xcolor}
\usepackage{relsize}

\UseRawInputEncoding

\usepackage{blindtext}

\usepackage{color}
\usepackage{tikz}
\usetikzlibrary{shapes}
\usepackage[normalem]{ulem}
\usepackage{mathtools}
\usepackage{fixltx2e}
\usepackage{color}
\usepackage{lineno} 
\usepackage{amsfonts}
\usepackage{soul}
\usepackage{bm}
\usepackage{float}

\begin{document}
    \title{Continuous Approximation of Stochastic Maps for Modeling Asymmetric Cell Division}
\author{Aviv Arcobi}
\email{avivarcobi@gmail.com}
\author{Stanislav Burov}
\email{stasbur@gmail.com}
\affiliation{Physics Department, Bar-Ilan University, Ramat Gan 5290002,
Israel}


\begin{abstract}
  Cell size control and homeostasis is a major topic in cell biology yet to be fully understood. 
Several growth laws like the timer, adder, and sizer were proposed, and mathematical approaches that model cell growth and division were developed. 
This study focuses on utilizing stochastic map modeling for investigating asymmetric cell division. 
We establish a mapping between the description of cell growth and division and the Ornstein-Uhlenbeck process with dichotomous noise. 
We leverage this mapping to achieve analytical solutions and derive a closed-form expression for the stable cell size distribution under asymmetric division. 
To validate our findings, we conduct numerical simulations encompassing several cell growth scenarios.
Our approach allows us to obtain a precise criterion for a bi-phasic behavior of the cell size.  
While for the case of the sizer scenario, a transition from the uni-modal phase to bi-modal is always possible, given sufficiently large asymmetry at the division, the affine-linear approximation of the adder scenario invariably yields uni-modal distribution.
\end{abstract}

\maketitle


\section{Introduction}
\label{chapter introductionNew}

The growth of individual bacteria reveals a remarkable pattern of exponential growth observed across various bacterial types~\cite{wang,mir}.
This exponential growth is characterized by the cell volume $v(t)$ following the equation $v(t)=e^{t/\tau}$, with $\tau$ denoting the average doubling time, until the occurrence of a division event. 
However, the intriguing question arises: how does the cell determine the right time for division?
Experimental observations shed light on this phenomenon by demonstrating the existence of a stable cell size distribution at the time of division, which strongly suggests the presence of a cell size regulation mechanism.
This mechanism involves a feedback system whereby bacteria actively control their size at division based on their size at birth, commonly referred to as a growth model or growth law~\cite{taheri,jun, howardsizer2017, Amir2018}

Early studies~\cite{koch1962} initially proposed the "sizer" model, suggesting that cells grow until reaching a critical size before undergoing division. 
Another prominent paradigm is the "timer" model, wherein cells aim to grow for a specific duration of time. Moreover, the suggested coupling between DNA replication and cell division led to the development of the "sizer+timer" model, in which the completion of a critical size threshold triggers division after a constant period of time~\cite{johnston,fantes}.
Alternative models have been proposed, including the ``adder" model~\cite{voorn,amir,osella,fantes}. 
According to the ``adder" model, cells consistently add a constant size between birth and division, irrespective of their size at birth. 
Numerous studies~\cite{amir,taheri,campos,deforet,jun,sauls} provide substantial evidence, confirming the validity of the ``adder" model as a mechanism responsible for maintaining size homeostasis in bacteria.

The mathematical modeling of bacteria growth laws has been the subject of numerous studies~\cite{Amir2018, Thomas2018, Grima2021, Fu2023, Chou2020, Chou2021}, focusing on the dynamic evolution of cell size following each division event. 
This evolution is typically represented by the relation between the cell size after division event number $n$, denoted as $v_n$, and the subsequent cell size after the next division event, represented as $v_{n+1}$. 
The developed equations strive to emulate and replicate the characteristics of diverse growth models, such as ``sizer," ``timer," and ``adder". 
By incorporating these mathematical formulations, researchers aim to uncover the underlying  mechanisms governing bacterial growth, providing valuable insights into the regulation of cell size in different biological contexts, like the bacteria population growth~\cite{Amir2017,barber2021}.

According to the mathematical description presented in \cite{amir}, the size of bacteria at the $(n+1)$st division event, denoted as $v_n e^{t_{n+1}/\tau}$, is governed by the growth law function $f(v_n)=v_n e^{t_{n+1}/\tau}$.
Here $t_{n+1}$ is the duration of the growth cycle between the $n$th and $(n+1)$st division events and is provided by $t_{n+1}=\tau\left[\ln\left(f(v_n)/v_n\right)\right]$. 
To account for stochastic elements in the growth process, a random noise term $\eta_n$ is introduced and assumed to be a Gaussian variable with zero mean.
In the case of symmetric division, where the cell divides into two halves, the size of the cell at the division event is given by
\begin{equation}
v_{n+1}=\frac{1}{2} v_{n} e^{\left[\ln{(f(v_{n})/v_{n})}+ \eta_{n}\right]}
    \label{eq:introductionGen}.
\end{equation}
In the case of the ``sizer" growth model, cell division occurs at a critical size, and the corresponding growth law is given by $f(v_n) = C$, where $C$ is a positive constant. 
In contrast, the ``timer" model assumes a constant growth period with $t_{n+1} = B\tau$, and the growth law is expressed as $f(v_n) = e^B v_n$ (where $B>0$). For the ``adder" model, the growth law is represented by $f(v_n) = v_n + A$, where a constant amount is added to the cell size. 
For the ``timer" model, Eq.~\eqref{eq:introductionGen} shows that no stable distribution of cell sizes exists. 
While for the ``adder" model, Eq.~\eqref{eq:introductionGen} suggests that the stable distribution of the cell sizes results in a log-normal distribution~\cite{amir} 

Bacteria can divide symmetrically, e.g. {\it Bascillus subtilis} and {\it Escherichia coli}, or asymmetrically, e.g., {\it Saccharomyces cerevisiae}~\cite{hartwell1977}, {\it Caulobacter crescentus}~\cite{skerker2004}, and asymmetrically dividing mycobacteria~\cite{logsdon2018}.  In this study, we tackle the issue of asymmetric division in the cell growth process.
Specifically, we investigate how Eq.\eqref{eq:introductionGen} is modified and explore the stable cell size distributions that arise when cell division is asymmetric~\cite{Huang2021, Grima2022,barber2021}. 
While previous research conducted by Marantan and Amir \cite{marantan&amir} has  examined this question, we propose a  novel approach that involves a continuous approximation of the discrete process described by Eq.\eqref{eq:introductionGen}.
We introduce a technique to map the problem of the stochastic map that defines the growth and division process to the Ornstein-Uhlenbeck process with dichotomous noise. 
This technique allows us to move beyond calculating only the moments of the cell size and derive an explicit analytic expression for the cell-size distribution.
In cases where the growth law $f(v)$ follows an affine linear relationship \cite{marantan&amir}, specifically $\ln(f(v)/v)=A-\alpha \ln(v_n)$, we present a detailed analysis of the cell-size distribution. We identify a transition between uni-modal and bi-modal phases and determine the precise expression that governs the phase transition point/line.
Additionally, we explore the scenario where the ``adder" growth law is approximated using the affine-linear model.
Our investigation demonstrates that the resultant cell-size distribution exclusively displays a uni-modal shape in such cases.
This observation underscores the potential limitation of the affine linear approximation when contrasted with the non-linear adder scenario. 
 Conversely, we observe a transition from uni-modal to bi-modal behavior for the sizer scenario. 
 Moreover, we present a simple formula for the phase transition boundary, which depends on the noise strength and asymmetry parameter.

The structure of this paper is as follows: In the Section ``Stochastic Maps and Langevin Equations," we present the continuous approximation approach for symmetric division and elucidate the methodology for constructing an analytical solution. Subsequently, we extend this framework to encompass asymmetric division and develop an analytic solution in Fourier and real space. We also explore the conditions governing the transition from uni-modal to bi-modal behaviors.
In the subsequent section, ``Comparison to Specific Growth Laws," we comprehensively compare the obtained analytic expressions with simulations conducted using the timer, adder, and sizer models.
A summary and discussion are provided in the section ``Summary".

\section{Stochastic Maps and Langevin Equations}
In this section, we present a detailed description of the continuous approximation method employed for equations similar to Eq.\eqref{eq:introductionGen}. We begin by outlining the approach initially developed in \cite{kessler&burov} for the symmetric division scenario, characterized by Eq.\eqref{eq:introductionGen}. Subsequently, we extend this method to address the  case of asymmetric division.

\subsection{Symmetric Division}

We introduce the variable $a_n = \ln(v_n)$ to establish a logarithmic transformation. Through this transformation, we can express Eq.~\eqref{eq:introductionGen} in an equivalent form:
\begin{equation}
\label{eq:SM a}
a_{n+1} = a_{n} + g(a_{n}) + \eta_{n}
\end{equation}
In this equation, $g(a_n) = \ln\left[f\left(e^{a_n}\right)\right] - a_n - \ln(2)$. The formulation presented in Eq.~\eqref{eq:SM a} represents a stochastic map, serving as a discrete counterpart to a stochastic differential equation. In the stochastic map framework, the discrete parameter $n$ assumes the role of continuous time $t$, allowing for the analysis of dynamic processes over discrete time intervals.
Stochastic maps find applications in diverse fields such as Mathematics \cite{kesten}, Finance \cite{follmer}, and Biology \cite{marantan&amir}. These maps provide a valuable tool for investigating the dynamics of complex systems characterized by randomness and discrete evolution. 

We have previously established that the growth law defines the function $f()$. In the subsequent analysis, we adopt the affine linear approximation originally introduced by \cite{amir}:
\begin{equation}
\label{eq:affine-linear}
f(v_{n}) = 2 C^{\alpha} v_{n}^{1-\alpha}
\end{equation}
It is worth noting that this approximation coincides with other growth models under specific parameter values. 
When $\alpha$ takes the values $\alpha = 0$ and $\alpha = 1$, the model perfectly aligns with the timer and sizer models, respectively. Furthermore, when $\alpha = \frac{1}{2}$, the $g(a_n)$ approximates the adder model to the lowest order of expansion around $a_n=\ln[A]$.
Therefore by tuning the parameter $0\leq\alpha\leq 1$, the affine linear approximation covers all three major growth models. 

In the work by Kessler and Burov \cite{kessler&burov},  a continuous approximation for the stochastic map  described by Eq.\eqref{eq:SM a} was introduced. 
This approximation is motivated by the resemblance of Eq.\eqref{eq:SM a} to the Euler-Maruyama discretization of a Langevin equation. Known as the second-order approximation for stochastic maps, as presented in \cite{kessler&burov}, it employs a continuous variable $t$ (time) instead of the generation number $n$. The corresponding Langevin equation for $f(v_n)$ (in  Eq.~\eqref{eq:affine-linear}) is expressed as:
\begin{equation}
\label{eq:Langevin symmetric}
da_{t}=\frac{1}{1-\frac{1}{2}\alpha} \left[-\alpha a_{t}+\alpha \ln(C)\right]dt +\frac{1}{1-\frac{1}{2}\alpha} \sqrt{\langle\eta^2\rangle}dB_{t}
\end{equation}
Here, $B_t$ represents the Wiener process, corresponding to the Langevin equation's noise term. 
Once the Langevin equation is derived, we can interpret the biological model from a physical standpoint. The cellular division and growth process can be conceptualized as akin to a particle immersed in a bath at a defined temperature (which corresponds to the diffusion term in the Langevin equation), while being subjected to the effects of an external potential (the drift term in the Langevin equation).
For Eq.~\eqref{eq:Langevin symmetric}, this Langevin equation describes the Ornstein-Uhlenbeck process~\cite{gardiner}.
The stable distribution of cell sizes is represented by the equilibrium distribution of the particle in the bath after a long time.

To determine this equilibrium distribution, we write the corresponding Fokker-Planck equation for $a_t$,
\begin{equation}
\label{eq:FP symmetric}
\frac{\partial{P(a,t)}}{\partial{t}} = T \frac{\partial^2 {P(a,t)}}{\partial{a^2}} +\frac{\partial}{\partial{a}}[U'(a) P(a,t)]
\end{equation}
Here, $P(a,t)$ represents the probability density function (PDF) of the random variable $a_t$. The term $T=\frac{1}{2} (\frac{1}{1-\frac{1}{2}\alpha})^2 \langle\eta^2\rangle$ serves as an analog of temperature, while  $U = \frac{1}{1-\frac{1}{2}\alpha} (\frac{\alpha}{2} a_{t}^2 -\alpha \ln({C}) a_{t})$ is the effective potential.
The stable distribution is obtained by taking the limit  $t\to\infty$ where $P(a,t)\underset{t\to\infty}{\longrightarrow}P_{sym}(a)$ and 
\begin{equation}
    \label{eq:sym solution}
    P_{sym}(a) =\mathcal{N}^{-1} e^{-\frac{U(a)}{T}}
\end{equation}
is the Maxwell-Boltzmann distribution where $\mathcal{N}=\int_{-\infty}^{\infty} exp[-\frac{U(a)}{T}] \,da\ $ is the normalization constant, and the subscript $sym$ means that this is the PDF for the case of symmetric division.

While here we have utilized a specific affine linear approximation, the form of Eq.~\eqref{eq:sym solution} is  general and applicable to any form of the growth law, $f(v_n)$. 
In the following section, we apply the described method of continuous approximation to the case of asymmetric division.

\subsection{Asymmetric Division}

The prefactor $\frac{1}{2}$ in Eq.~\eqref{eq:introductionGen} signifies that during division, the cell is split into two halves, and $v_{n+1}$ represents one of the offspring. 
Since the division is symmetric, it is not essential which offspring is described by $v_{n+1}$.
However, in the case of asymmetric division where the two offspring differ in size, Eq.~\eqref{eq:introductionGen} is transformed into:
\begin{equation}
\label{eq:SM v asymmetric}
v_{n+1} = \Delta_{\pm} v_{n} \exp\left[\ln\left({f(v_{n})}/{v_{n}}\right) + \eta_{n}\right]
\end{equation}
Here, $\Delta_+$ is used if we select the larger offspring for $v_{n+1}$, and $\Delta_-$ is used if the smaller one is selected. 
The ratios $\Delta_\pm$ satisfy $\Delta_+ + \Delta_- = 1$, and we assume that they are independent of $n$. 

How we decide which offspring, or $\Delta_\pm$, to choose? This decision typically depends on the experimental protocol. For instance, in \cite{sriIyerBiswas}, only the larger offspring, referred to as the ``mother," is kept. In such a scenario, the treatment of stochastic models describing the evolution as a function of $n$ is very similar to the symmetric case, with the only difference being the replacement of the prefactor $1/2$ in Eq.~\eqref{eq:introductionGen} with a different constant.
This work considers the situation where all the offspring are kept. The question regarding the stable cell size distribution is addressed as follows: after numerous generations since the process started, we investigate the probability of observing a cell of size $v$ when randomly choosing one of the many offspring constituting the cell colony.
 This random selection of a cell is equivalent to randomly choosing a timeline, starting from the original cell and making random choices of which offspring to follow at each division event. We assume the absence of bias, meaning that we choose $\Delta_+$ and $\Delta_-$ with equal probability of $1/2$.

To summarize, in the case of asymmetric division, we utilize Eq.\eqref{eq:SM v asymmetric}, and for each value of $n$, we randomly determine the choice of $\Delta_\pm$ by ``tossing a coin." 
Therefore, in Eq.\eqref{eq:SM v asymmetric}, two stochastic terms are present: (i) $\eta_n$, which models the stochastic variations during the growth process, and (ii) $\Delta_{\pm}$, which determines the random choices during the selection of timelines. 
We assume the validity of the affine linear approximation for $f(v_n)$ as described in Eq.~\eqref{eq:affine-linear}.

The variable $a_n = \ln(v_n)$ satisfies Eq.\eqref{eq:SM a} with the function
\begin{equation}
\label{eq:gn asym definition}
g(a_{n}) = \alpha \ln({C}) - \alpha a_{n} + \ln(2) + \ln(\Delta_{\pm}).
\end{equation}
Applying the second-order continuous approximation protocol, similar to the symmetric division case, we derive the Langevin equation
\begin{equation}
\label{eq:Langevin asymmetric}
da_{t} = \frac{-\alpha a_{t} + \alpha \ln({C}) + \ln(2) + \ln(\Delta_{\pm})}{1-\frac{1}{2}\alpha} dt + \frac{\sqrt{\langle\eta^2\rangle}}{1-\frac{1}{2}\alpha} dB_{t}.
\end{equation}
It is worth noting that the term in Eq.\eqref{eq:Langevin asymmetric} is time-dependent. 
According to the protocol, the value of $\Delta_\pm$ is randomly chosen at $t = n = 1, 2, 3, \ldots$.
For any non-integer $t$, $\Delta_\pm$ takes the value it obtained at the nearest integer $n \leq t$. Consequently, $\Delta_\pm$ is a dichotomous random process that can switch between $\Delta_+$ and $\Delta_-$ at integer values of $t$.
If $\Delta_\pm$ switched  to $\Delta_-$/$\Delta_+$ at $t=n_1$, the amount of time $n_2-n_1$ that will pass before $\Delta_\pm$ switches back to  $\Delta_+$/$\Delta_-$ is geometrically distributed, i.e. $p(n_2-n_1)=(1/2) (1-1/2)^{n_2-n_1}$, where $n_2-n_1=1,2,3,...$. By using $\tau_\pm=n_2-n_1$ as the waiting time in the $+$ or the $-$ state, we approximate the discrete geometrical  distribution of $\tau_\pm$ by its continuous analog, i.e., the exponential distribution 
\begin{equation}
    \label{eq:dichotomous time}
p(\tau_\pm)=\gamma e^{-\gamma \tau_\pm} 
\end{equation}
where $\gamma$ is the switching rate. 
To fit the discrete process $\gamma$ must satisfy $\gamma=-\ln(1-1/2)$.
Further on we use a continuous representation for $\Delta_\pm$: a dichotomous noise that switches between $\Delta_+$ and $\Delta_-$ at a rate $\gamma$. 
Dichotomous noise, as described in \cite{bena}, finds widespread application in modeling diverse phenomena across Physics \cite{doering}, Biology \cite{mankin}, and Chemistry \cite{reimann}.  

While the Gaussian noise in the Langevin equation for $a_t$, Eq.\eqref{eq:Langevin asymmetric}, resembles the noise in the symmetric case, the presence of dichotomous noise introduces two distinct states for $a_t$. One state, denoted as $+$, corresponds to $\Delta_\pm = \Delta_+$, while the other state, denoted as $-$, corresponds to $\Delta_\pm = \Delta_-$. To handle the dichotomous noise, we rewrite the term $\ln(\Delta_\pm)$ symmetrically as $\ln(\Delta_\pm) = [\ln(\Delta_+) + \ln(\Delta_-)]/2 \pm \delta$. By defining $x_t = a_t - \ln(C) - \ln(2)/\alpha - [\ln(\Delta_+) + \ln(\Delta_-)]/2\alpha$, we can transform Eq.\eqref{eq:Langevin asymmetric} into the following form:
\begin{equation}
    \label{eq:xt Langevin}
    dx_t=\left(-{\tilde \alpha}x_t\pm{\tilde \delta}\right)\,dt + \sqrt{2T} dB_t
\end{equation}
where ${\tilde \alpha} = \alpha/(1-\alpha/2)$, ${\tilde \delta} = \delta/(1-\alpha/2)$, and $T=\frac{1}{2}(\frac{1}{1-\frac{1}{2}\alpha})^2\langle\eta^2\rangle$ serves as an analog for temperature. The probabilities of being in state $+$ or $-$ are denoted as $P_+(x,t)$ and $P_-(x,t)$, respectively. The Langevin equation \eqref{eq:xt Langevin} leads to a pair of coupled Fokker-Planck equations for $P_+(x,t)$ and $P_-(x,t)$ \cite{GittermanBurov,bena,fang2022}:
\begin{equation}
    \label{eq:coupled fokkerP}
    \begin{array}{l}
         \frac{\partial{P_{+}(x,t)}}{\partial{t}} = 
         T \frac{\partial^2 {P_{+}(x,t)}}{\partial{x^2}} +
         \\
         \frac{\partial}{\partial{x}}\left[\left(U'(x)+{\tilde \delta}\right) P_{+}(x,t)\right] -\gamma P_{+}(x,t) + \gamma P_{-}(x,t)
           \\
          \frac{\partial{P_{-}(x,t)}}{\partial{t}} = 
          T \frac{\partial^2 {P_{-}(x,t)}}{\partial{x^2}} +
          \\
          \frac{\partial}{\partial{x}}\left[\left(U'(x)-{\tilde \delta}\right) P_{-}(x,t)\right] -\gamma P_{-}(x,t) + \gamma P_{+}(x,t) ,
          
    \end{array}
\end{equation}
where 
\begin{equation}
    \label{eq:potential definition}
    U(x) = \frac{1}{2}{\tilde \alpha}x^2
\end{equation}
and $U'(x) = dU(x)/dx$. 
Equations \eqref{eq:coupled fokkerP} and \eqref{eq:potential definition} establish a mapping from the model of cell size with asymmetric division to the Ornstein-Uhlenbeck process with dichotomous noise.
This mapping corresponds to the problem of a particle in a harmonic potential where the minima of the potential fluctuate between two distinct values, denoted as $x_m = \pm{\tilde \delta}$.
We have established that  akin to the symmetric case, cell division and growth can be mapped to behavior of a particle in a potential landscape, coupled to a heat bath at temperature $T$. 
However, in the presence of division asymmetry, an additional noise term arises, resulting in random fluctuations of the potential. 
The problem of finding the stable size distribution reduces to finding the positional PDF of a particle in a fluctuating potential landscape~\cite{dybiec,schwarcz2019}.

We take the limit $t\to\infty$ and assume that the process reaches a steady state. Then the terms $\partial P_{\pm}(x,t)/\partial t$ in Eq.~\eqref{eq:coupled fokkerP} become negligible. As a result, the coupled equations can be decoupled as detailed in Appendix A and \cite{dybiec}. 
The PDF of finding the process at $x$, irrespective of $+$ or $-$ state,  $P(x)=\left(P_+(x)+P_-(x)\right)/2$ is provided by
\begin{equation}
\label{eq:3rd order ode}
    \begin{array}{l}
    T^2 P'''(x) + 2TU'(x)P''(x)+
    \\
     \left[ 3 T U''(x)+U'(x)^2-\tilde{\delta}^{2} -2 \gamma T\right] P'(x)+ 
     \\ 
     \left[ 2 U'(x)U''(x) + T U'''(x) -2\gamma U'(x)\right] P(x)=0, 
    \end{array}
\end{equation} 
which is a linear ordinary differential equation of third order.
Below we present an explicit solution to Eq.\eqref{eq:3rd order ode} for the case where the potential $U(x)$ takes on a harmonic form, corresponding to the affine linear approximation of the growth law described by Eq.\eqref{eq:affine-linear}.

\subsection{Solution of the Continuous Approximation}

While Equation~\eqref{eq:3rd order ode} has been previously solved in \cite{dybiec} for specific cases with $T=0$ or $\gamma=0$, in this section we will derive an explicit solution for the general case. 
The solution method involves a transformation to Fourier space, a solution of a second-order ordinary differential equation in $k$-space followed by inversion back to $x$ space, power series expansion, and partial summation. To  bypass intricate technicalities, readers can proceed directly to Equation~\eqref{eq:final x series expansion} for the final formula and the following discussion.  

Transformation of Eq.~\eqref{eq:3rd order ode} into Fourier space results in
\begin{equation}
\label{eq:2nd order ode k}
\begin{array}{l}
   \tilde{\alpha}^2  k \hat{P}''(k) +  \left[ 2 T \tilde{\alpha} k^2 +2 \gamma \tilde{\alpha}\right] \hat{P}'(k) +
   \\ \left[ T^2 k^3 +  T \tilde{\alpha} k +  \tilde{\delta}^2 k + 2 \gamma T k \right] \hat{P}(k) = 0
\end{array}
\end{equation} 
where ${\hat P}(k)=\int_{-\infty}^{\infty}e^{-ikx}P(x)\,dx$. 
 Ordinary differential equation similar to Eq.~\eqref{eq:2nd order ode k} appears at \cite{kamke}, and we follow the course of action presented there. 
 First, we substitute $\hat{P}(k)=e^{-\frac{T k^2}{2 \tilde{\alpha}}} u(k)$ and obtain
\begin{equation}
\label{eq:ode u}
    \tilde{\alpha}^2 k u''(k)+2 \gamma \tilde{\alpha} u'(k)+\tilde{\delta}^2 k u(k)=0
\end{equation}
then we assign $u(k)=k^{\frac12 -\frac{\gamma}{\tilde{\alpha}}} y(k)$ and $z=\frac{\tilde{\delta}}{\tilde{\alpha}} k$ , that leads  to the Bessel Equation
\begin{equation}
\label{eq:bessel ode}
   z^2 y''(z)+ z y'(z) +[z^2 -( \frac{\gamma}{\tilde{\alpha}}-\frac12)^2] y(z)=0
\end{equation}
with a general solution of the form 
\begin{equation}
\label{eq:bessel solution}
    y(z)=B_1 J_{\frac{\gamma}{\tilde{\alpha}}-\frac12}\left( z\right)+ B_2 Y_{\frac{\gamma}{\tilde{\alpha}}-\frac12}\left( z\right)
\end{equation}
where $J_w(\dots)$ and $Y_w(\dots)$ are the Bessel functions of the first and second kind, respectively, and $B_1$, and $B_2$ are constants that yet to be determined. 
Therefore, the general solution of Eq. \eqref{eq:2nd order ode k} is
\begin{equation}
\label{eq:general solution k}
\begin{array}{l}
    \hat{P}(k)= e^{-\frac{k^2 T}{2 \tilde{\alpha}}} k^{\frac{1}{2} - \frac{\gamma}{\tilde{\alpha}}} 
    \times
    \\
    \left[B_1 J_{ \frac{\gamma}{\tilde{\alpha}}-\frac12}\left( \frac{\tilde \delta k}{\tilde \alpha}\right)+ B_2 Y_{ \frac{\gamma}{\tilde{\alpha}}-\frac12}\left( \frac{\tilde \delta k}{\tilde \alpha}\right)\right].
    \end{array}
\end{equation}
In order to find the constants $B_1$ and $B_2$ we recall that
\begin{equation}
\label{eq:moment-generating function}
    \langle x^n\rangle=(-i)^n \frac{d^n}{d k^n}\hat{P}(k=0).  
\end{equation}
Expansion around $k=0$ of the Bessel function of the second kind $Y$ produces powers of $k$ in the form of $k^{1-\frac{2\gamma}{\tilde{\alpha}}}\sum_{n=0}^{\infty} k^{2 n}$ and according to Eq.~\eqref{eq:moment-generating function} this leads to diverging moments of $x$. Therefore we set $B_2=0$ and by utilizing the normalization condition for $P(x)$, i.e. ${\hat P}(k=0)=1$ we obtain that
$B_1=\Gamma[\frac{1}{2} + \frac{\gamma}{\tilde{\alpha}}] (\frac{\tilde \delta}{2 \tilde\alpha})^{\frac{1}{2}-\frac{\gamma}{\tilde{\alpha}}}$. 
Finally,
\begin{equation}
\label{eq:k solution with constants}
    \hat{P}(k)=\Gamma[\frac{1}{2} + \frac{\gamma}{\tilde{\alpha}}] (\frac{{\tilde\delta} k}{2 \tilde\alpha})^{\frac{1}{2}-\frac{\gamma}{\tilde{\alpha}}} e^{-\frac{k^2 T}{2 \tilde{\alpha}}} J_{ \frac{\gamma}{\tilde{\alpha}}-\frac12}\left( \frac{\tilde \delta k}{\tilde \alpha}\right)
\end{equation}
is the form of $P(x)$ in the Fourier space.
To obtain the solution in $x$ space we use  the series expansion of $J_w(y)$: 
 $   J_w(y)=\sum_{m=0}^{\infty} \left[(-1)^m/(m! \Gamma[m+w+1]) \right](y/2)^{2m+w}\
 $
\cite{abramowitz}
and the fact that $(-k^2)^m e^{-\frac{k^2 T}{2 \tilde{\alpha}}}=\frac{d^m}{d(\frac{T}{2 \tilde{\alpha}})^m} e^{-\frac{k^2 T}{2 \tilde{\alpha}}}$,
that leads to
\begin{equation}
\label{eq:final k solution}
    \hat{P}(k)= \Gamma[\frac{1}{2} + \frac{\gamma}{\tilde{\alpha}}] \sum_{m=0}^{\infty} \frac{\left(\frac{\tilde\delta}{2 \tilde\alpha}\right)^{2m}}{m! \Gamma[m+\frac{1}{2} + \frac{\gamma}{\tilde{\alpha}}]}  \frac{d^m}{d(\frac{T}{2 \tilde{\alpha}})^m} e^{-\frac{k^2 T}{2 \tilde{\alpha}}} .
\end{equation}
The Gaussian is easily inverted and finally the form of the PDF $P(x)$ is provided by
\begin{equation}
\label{eq:final x solution}
    {P}(x)= \frac{\Gamma[\frac{1}{2} + \frac{\gamma}{\tilde{\alpha}}]}{\sqrt{2 \pi}} \sum_{m=0}^{\infty} \frac{\left(\frac{\tilde\delta}{2 \tilde\alpha}\right)^{2m}}{m! \Gamma[m+\frac{1}{2} + \frac{\gamma}{\tilde{\alpha}}]}  \frac{d^m}{d(\tilde{T})^m} \frac {e^{-\frac{x^2}{4 \tilde{T}}}}{\sqrt{2 \tilde{T}}} 
\end{equation}  
where  $\tilde{T}=\frac{T}{2 \tilde{\alpha}}$.

While Eq.~\eqref{eq:final x solution} tells us that the solution is an infinite sum of Gaussians multiplied by polynomials of $x$, it is much more useful to present the power-series expansion of $P(x)$. 
To achieve this, we use Taylor expansion of the Gaussian function $e^{-x^2/4{\tilde T}}=\sum_{l=0}^\infty(-x^2/4{\tilde T})^l/l!$ and notice that $\frac{d^m}{d{\tilde T}^m}{\tilde T}^{-\frac{1}{2}-l}={\tilde T}^{-1/2-l-m}\Pi_{s=0}^{m-1}(-\frac{1}{2}-l-s)$. 
The product can be represented by Pochhammer symbol $(b)_m=b(b+1)\dots(b+m-1)=\Gamma[b+m]/\Gamma[b]$~\cite{abramowitz}, and therefore 
$\frac{d^m}{d{\tilde T}^m}{\tilde T}^{-\frac{1}{2}-l}={\tilde T}^{-1/2-l-m}\Gamma[\frac{1}{2}-l]/\Gamma[\frac{1}{2}-l-m]$. 
Since both $m$ and $l$ are integers, the $\Gamma$ function satisfies $\Gamma[\frac{1}{2}-l-m]=(-1)^{m-1}\Gamma[l-\frac{1}{2}]\Gamma[\frac{3}{2}-l]/\Gamma[\frac{1}{2}+l+m]$. We rewrite Eq.~\eqref{eq:final x solution} as
\begin{equation}
    \label{eq:semi-final x series}
    P(x)=\frac{1}{\sqrt{4\pi{\tilde T}}}\sum_{l=0}^\infty\frac{\left(-\frac{x^2}{4{\tilde T}}\right)^{l}}{l!}
    \sum_{m=0}^\infty \frac{\frac{\Gamma[\frac{1}{2}+l+m]}{\Gamma[\frac{1}{2}+l]}}{m!\frac{\Gamma[\frac{1}{2} + \frac{\gamma}{\tilde{\alpha}}+m]}{\Gamma[\frac{1}{2} + \frac{\gamma}{\tilde{\alpha}}]}}\left(-\frac{{\tilde\delta}^2}{4{\tilde T}{\tilde\alpha}^2}\right)^m
\end{equation}
and since $\sum_{m=0}^\infty z^m (b)_m/(c)_m=\,_1F_1[b,c;z]$ is the Kummer function~\cite{abramowitz} the power series expansion for $P(x)$  is
\begin{equation}
    \label{eq:final x series expansion}
    P(x)=\frac{1}{\sqrt{4\pi{\tilde T}}}\sum_{l=0}^\infty
   \frac{\,
    _1F_1\left[\frac{1}{2}+l,\frac{1}{2}+\frac{\gamma}{\tilde \alpha},-\frac{{\tilde\delta}^2}{4{\tilde T}{\tilde\alpha}^2}\right] }{l!}
    \left(-\frac{x^2}{4{\tilde T}}\right)^{l}.
\end{equation}
The Kummer function $_1F_1\left[\frac{1}{2}+l,\frac{1}{2}+\frac{\gamma}{\tilde \alpha},-\frac{\delta^2}{4{\tilde T}\alpha^2}\right]$ decays to zero as $l$ tends to infinity. With the presence of $l!$ in the denominator, this implies an infinite radius of convergence of the power-series in Eq.~\eqref{eq:final x series expansion}.

\begin{figure}[t]
    \centering
    \includegraphics[scale=0.5]{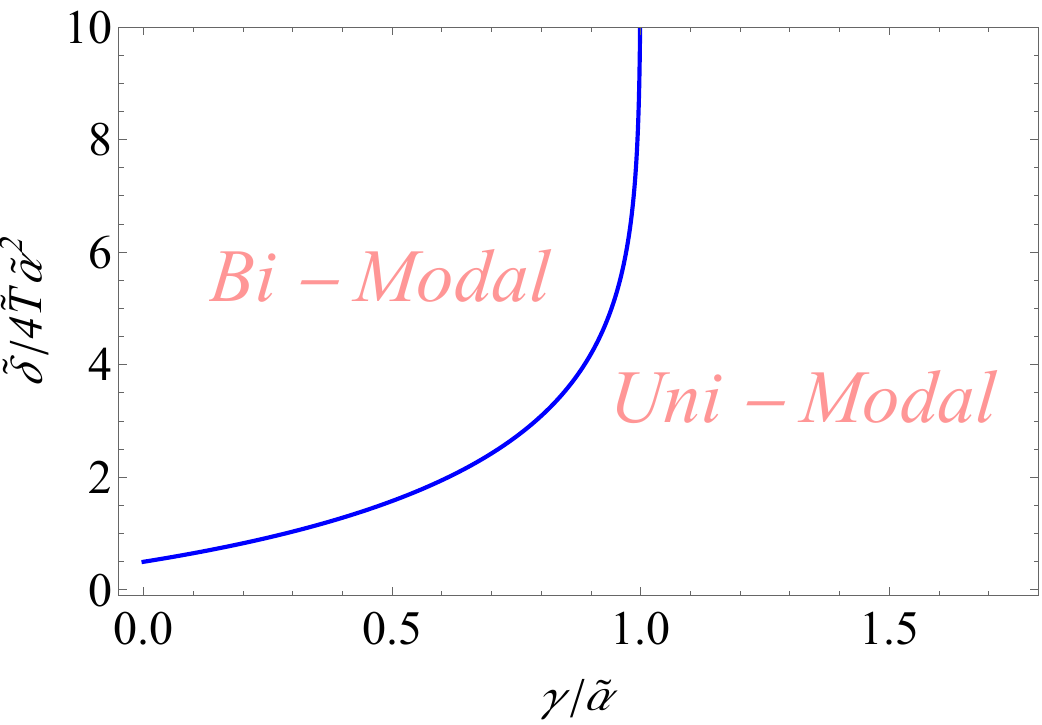}
    \caption{Phase diagram of a system described by the Langevin equation (Eq.~\eqref{eq:xt Langevin}), which models a Brownian particle moving in a harmonic potential $U= \frac{\tilde{\alpha}}{2} x_{t}^2$. The particle is subjected to a dichotomous noise with a strength of $\tilde{\delta}$ and a rate of $\gamma$.
The phase separation line, shown in blue, is determined by Eq.~\eqref{eq:phase separation}.  For values of $\gamma$ greater than $\tilde{\alpha}$, only the uni-modal phase exists.}
    \label{fig:phase-line}
\end{figure}

\begin{figure*}[bt!]
    \centering
    \includegraphics[trim = {4cm 9cm 4cm 9cm},scale=0.39]{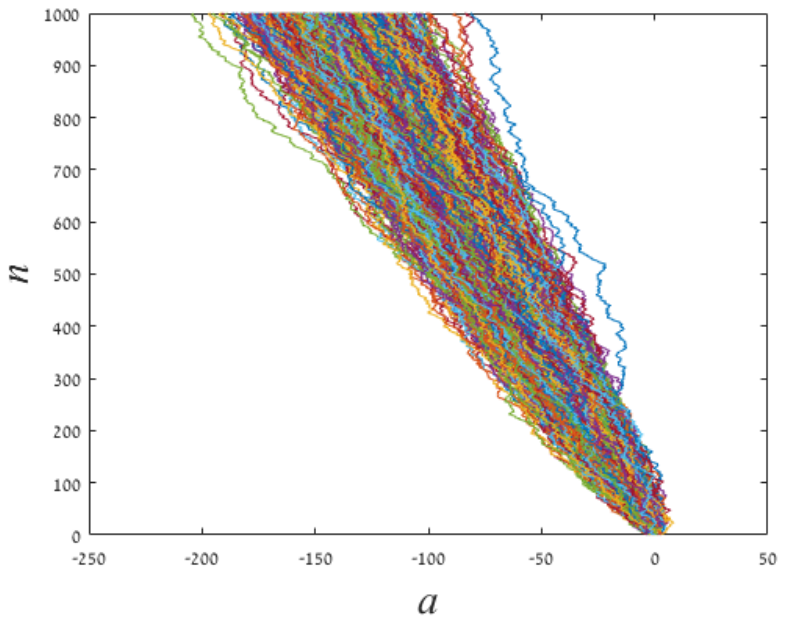}
    \hspace{2em}
    \includegraphics[trim = {4cm 9cm 4cm 9cm},scale=0.39]{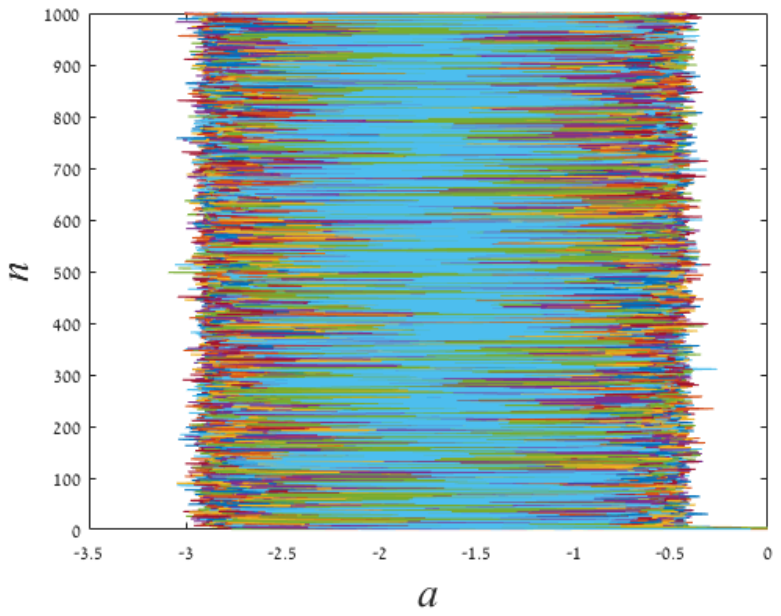}
     \hspace{2em}
    \includegraphics[trim = {4cm 9cm 4cm 9cm},scale=0.39]{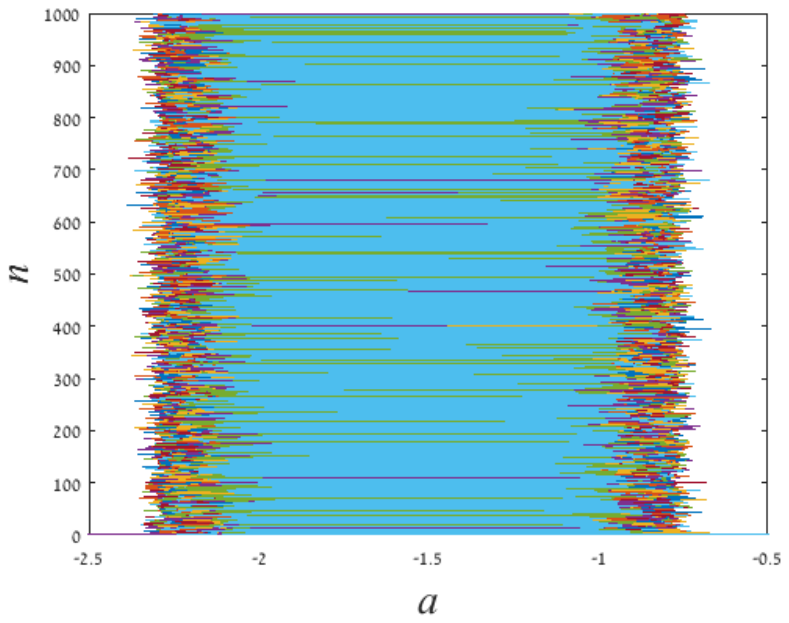}
    \caption{Numerical simulation of $1000$ paths of the stochastic process in  Eq.~\eqref{eq:SM a}  with $g(a_n)$ provided by Eq.~\eqref{eq:gn asym definition} and: $C=\frac14$, $\Delta_{+}=\frac34$, $\Delta_{-}=\frac14$, $\eta=\sqrt{0.005}$. {\bf Left panel:} The timer scenario, $\alpha=0$, all the paths diverge towards $-\infty$ hence it is unstable. {\bf Middle panel:} The adder scenario, $\alpha=0.5$ and {\bf Right panel:} The sizer scenario, $\alpha=1$. For both adder and sizer the paths are bounded and $a$ attain a stable distribution. Notice that, for the above parameters, the adder model is in a uni-modal state and the sizer is in a bi-modal state, therefor the observable distinction of two preferable states for the case of the sizer. }
    \label{fig:timer-example}
\end{figure*}

There are several notable observations regarding Eq.~\eqref{eq:final x series expansion}.
Firstly, when $\delta=0$, representing the symmetric division case, the Kummer function $_1F_1\left[\frac{1}{2}+l,\frac{1}{2}+\frac{\gamma}{\tilde \alpha},0\right]$ evaluates to 1 for all values of $l$. Consequently, the sum in Eq.~\eqref{eq:final x series expansion} becomes a Taylor expansion of a Gaussian function. This behavior is expected in the context of symmetric division since the affine-linear approximation of the growth law maps it to a particle in a harmonic potential.

Secondly, the expansion exclusively involves even powers of $x$, indicating that the function $P(x)$ reaches either a maximum or a minimum at $x=0$.
The transition between the maximum and minimum occurs when the coefficient of $x^2$ becomes zero. 
This transition signifies the existence of a unimodal phase and a bimodal phase. The phase separation line, which determines the transition between these phases, is determined by the zeros of the Kummer function:
\begin{equation}
\label{eq:phase separation}
_1F_1\left[\frac{3}{2},\frac{1}{2}+\frac{\gamma}{\tilde \alpha},-\frac{{\tilde\delta}^2}{4{\tilde T}{\tilde\alpha}^2}\right]=0
\end{equation}
Eq.\eqref{eq:phase separation} provides the phase separation line for an overdamped Brownian particle in a harmonic potential (Ornstein-Uhlenbeck process) subjected to dichotomous noise. 
The existence of uni-modal to bi-modal transition was demonstrated in\cite{dybiec}, but an explicit form of the phase separation line was previously only known for specific values of $T$ and $\gamma$.

In Fig.\ref{fig:phase-line}, the phase diagram is depicted, and the phase line (shown in blue) was obtained by numerically solving Eq.\eqref{eq:phase separation}. Notably, as $\gamma$ approaches $\tilde{\alpha}$, the phase separation line appears to diverge. 
This divergence can be understood through the integral representation of the Kummer function, $_1F_1[b,c,-z]=\Gamma[c]/\Gamma[b]\Gamma[c-b]\int_0^1e^{-zu}u^{b-1}(1-u)^{c-b-1},du$, for  $c > b$. Therefore $_1F_1[b,c,-z]$ only attains positive values when $c>b$. 
Henceforth, the coefficient of $-x^2/4{\tilde T}$ consistently holds a positive value, thereby signifying the presence of solely a uni-modal phase. 
Thus, there is a critical rate $\gamma_c$ of the dichotomous noise:
\begin{equation}
\label{eq:critical gamma}
\gamma_c = \tilde{\alpha}
\end{equation}
The critical rate $\gamma_c$ depends solely on the strength of the binding harmonic potential $\tilde{\alpha}$ and not on the noise strength $\delta$ or temperature $T$. 
When the jumps of the dichotomous noise occur too frequently, they prevent the process from stabilizing around the potential minima, even when these minima are widely separated.

Our discussion began with the introduction of a stochastic map for $a_n$, as defined in Eq.\eqref{eq:SM a}, considering an affine linear approximation for the growth law and asymmetric division. 
In the following section, we will examine the strengths and limitations of this approximation in the context of the timer, adder, and sizer models. 
However, before proceeding, it is beneficial to rewrite the stable distribution, given by Eq.\eqref{eq:final x series expansion}, in terms of the variable $a$ of the stochastic map and the original parameters $\alpha$, $\Delta_{\pm}$, $C$, and $\langle \eta^2 \rangle$.
The rewritten form of the stable distribution, denoted as $P(a)$, is given by:
\begin{widetext}
\begin{equation}
\label{eq:a stable distribution}
P(a) = \sqrt{\frac{\alpha(1-\frac{\alpha}{2})}{\langle \eta^2 \rangle\pi}} \mathlarger{\mathlarger{\mathlarger{\sum_{l=0}^\infty} }}\frac{1}{l!} \,
_1F_1\left[\frac{1}{2}+l,\frac{1}{2}+\frac{(1-\frac{\alpha}{2})\ln(2)}{\alpha};-\frac{(1-\frac{\alpha}{2})\ln\left(\sqrt{\frac{\Delta+}{\Delta_-}}\right)^2}{\alpha \langle \eta^2 \rangle}\right]  \left(-\frac{\left(a - \ln\left[C2^{\frac{1}{\alpha}}(\Delta_+\Delta_-)^{\frac{1}{2\alpha}}\right]\right)^2}{\langle \eta^2\rangle/ \alpha(1-\frac{\alpha}{2})}\right)^{l}.
\end{equation}
\end{widetext}
The stable PDF of the cell size $v$ therefore attains the form
\begin{equation}
\label{eq:v stable distribution}
{\tilde P}_{asym}(v) = \frac{1}{v}P(\ln(v)).
\end{equation}
It is important to note that in the above equation, we explicitly used $\gamma=\ln(2)$, corresponding to the random choice of offspring protocol as explained earlier. 
For different protocols of offspring selection, the value of $\gamma$ may vary.

\section{Comparison to Specific Growth Laws}

In this section, we will explore three widely used growth models: timer, adder, and sizer. Our focus will be on examining the performance of the developed continuous time approximation in comparison to the description provided by stochastic maps.

\subsection{Timer Model}

The timer model, as previously mentioned, is defined as a growth low that sets the growth time to a constant (up to stochastic fluctuations). It was found, quite a while ago~\cite{John1983,marantan&amir,taheri}, that this model can't lead to a stable size distribution when symmetric division is in place. In Fig.~\ref{fig:timer-example} $1000$, random time-lines of $a=\log(v)$ that follow the asymmetric timer growth law are plotted. The process doesn't seem to converge to a stable distribution but spreads mainly over the negative values of $a$.

The mapping from $a_t$ to $x_t$ involves dividing by $\alpha$, which in the case of the timer model is $0$. Consequently, we utilize Eq.~\eqref{eq:Langevin asymmetric} for $a$ with $\alpha=0$, and the mapping to $x$ is achieved by employing the potential $U(x)=-\left(\ln(2) + \left[\ln(\Delta_+\Delta_-)\right]/2\right)x$. Here, $dx_t=-\left(U'(x)\pm\delta\right)\,dt+\sqrt{2T}dB_t$. As a result, Eq. \eqref{eq:3rd order ode} simplifies to $T^2P'''(x)+2TU'(x)P''(x)+\left[U'(x)^2-\delta^2-2\gamma T\right]P'(x)-2\gamma U'(x)P(x)=0$, representing a linear ordinary differential equation with constant coefficients. The solution takes the form $P(x)=\sum_{j=1}^3 b_je^{r_j x}+c_j x e^{r_j x}+d_j x^2 e^{r_j x}$, where $b_j,c_j,d_j$ are constants. However, due to the divergence of $e^x$, this solution cannot be normalized. Consequently, there is no stationary solution for the timer model, even when accounting for the asymmetry of the division.

The spread observed in Fig.~\ref{fig:timer-example} can be easily understood by examining the behavior of $g(a_n)$. When $\alpha=0$, it becomes a negative constant with random fluctuations $\pm\delta$. This process resembles a discrete analog of a constant (negative) drift with random fluctuations. Thus, the inability of the timer model to converge to a stable distribution is reproduced by the non-existence of a normalizable $P(x)$.

\begin{figure}[t!]
    \centering
    \includegraphics[scale=0.42]{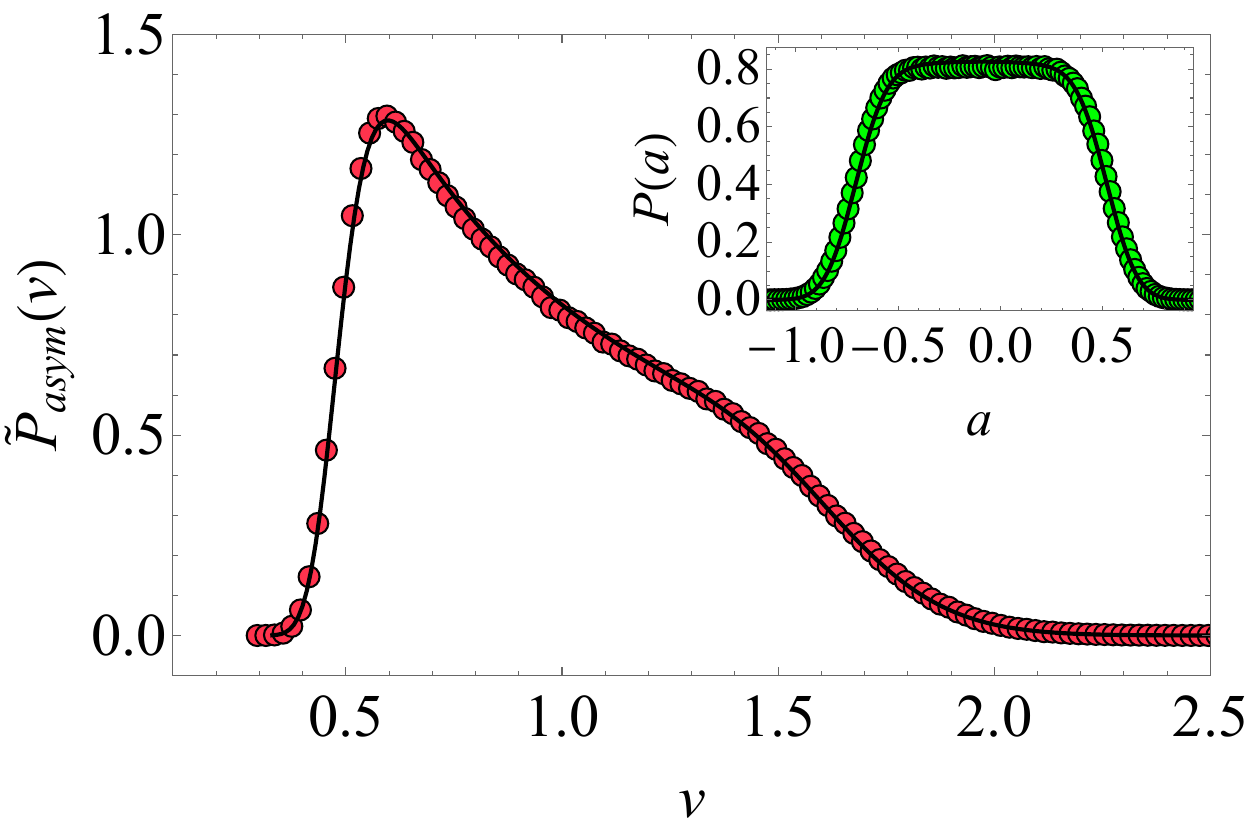}
    \\
    \caption{The stable distribution for adder growth law scenario, ${\tilde P}_{asym}(v)$ with affine - linear approximation and asymmetric division. Parameters : $\Delta_+=0.65$, $\alpha=1/2$, $\eta=0.1$, $C=1$, $10^7$ generations sampled. The inset is the distribution of $a=\ln(v)$. Symbols are the results of simulations while the thick line provided by Eq.~\eqref{eq:v stable distribution} for the figure and Eq.~\eqref{eq:a stable distribution} for the inset.
    }
    \label{fig:adder linear comparisom}
\end{figure}

\subsection{Adder Model}

\begin{figure*}[ht!]
\includegraphics[width=0.32\textwidth]{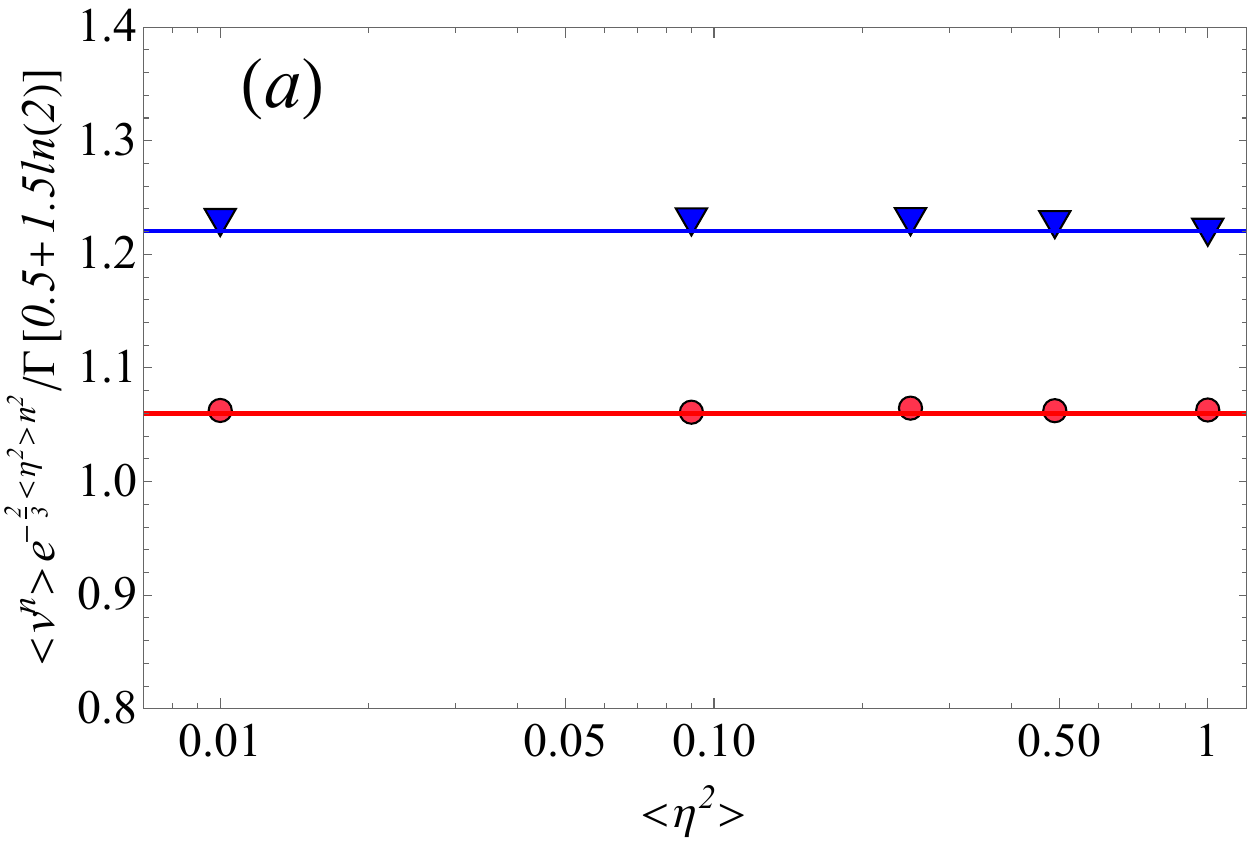} \hspace{0.2em}
\includegraphics[width=0.33\textwidth]{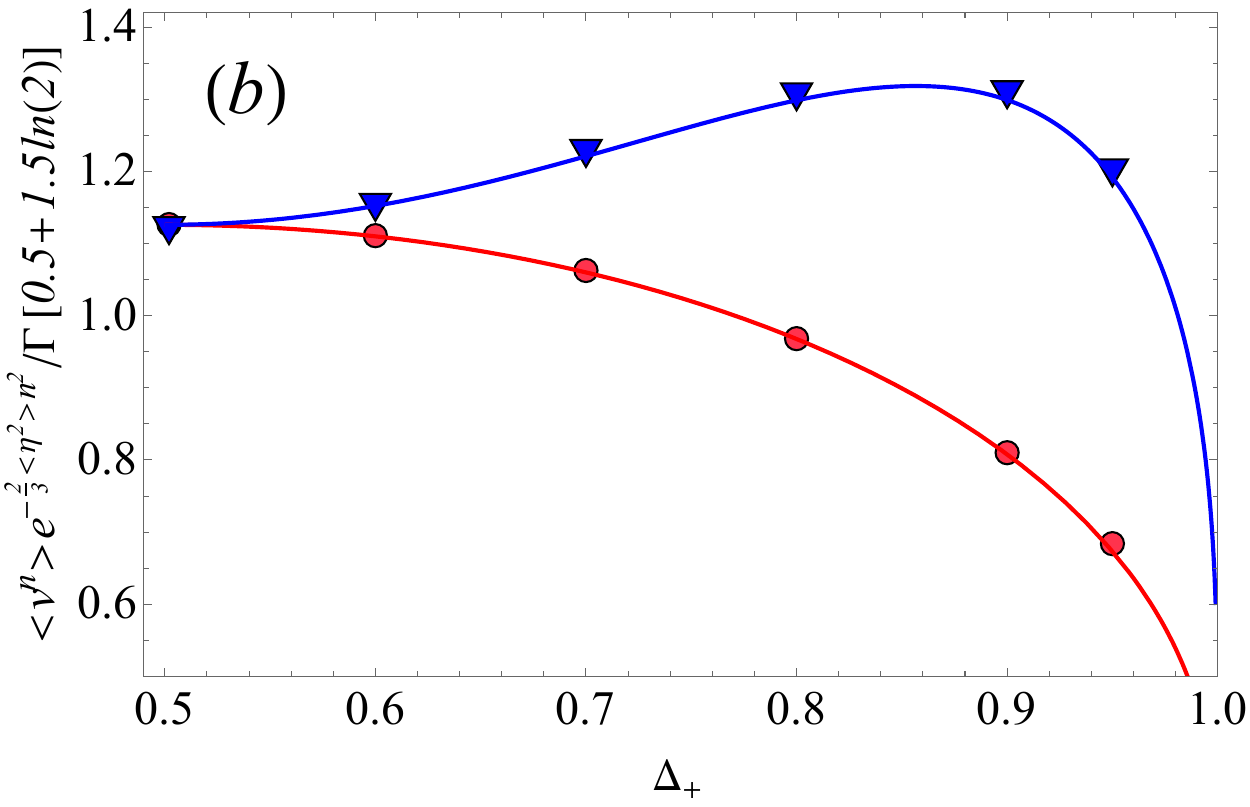} \hspace{0.2em}
    \includegraphics[width=0.32\textwidth]{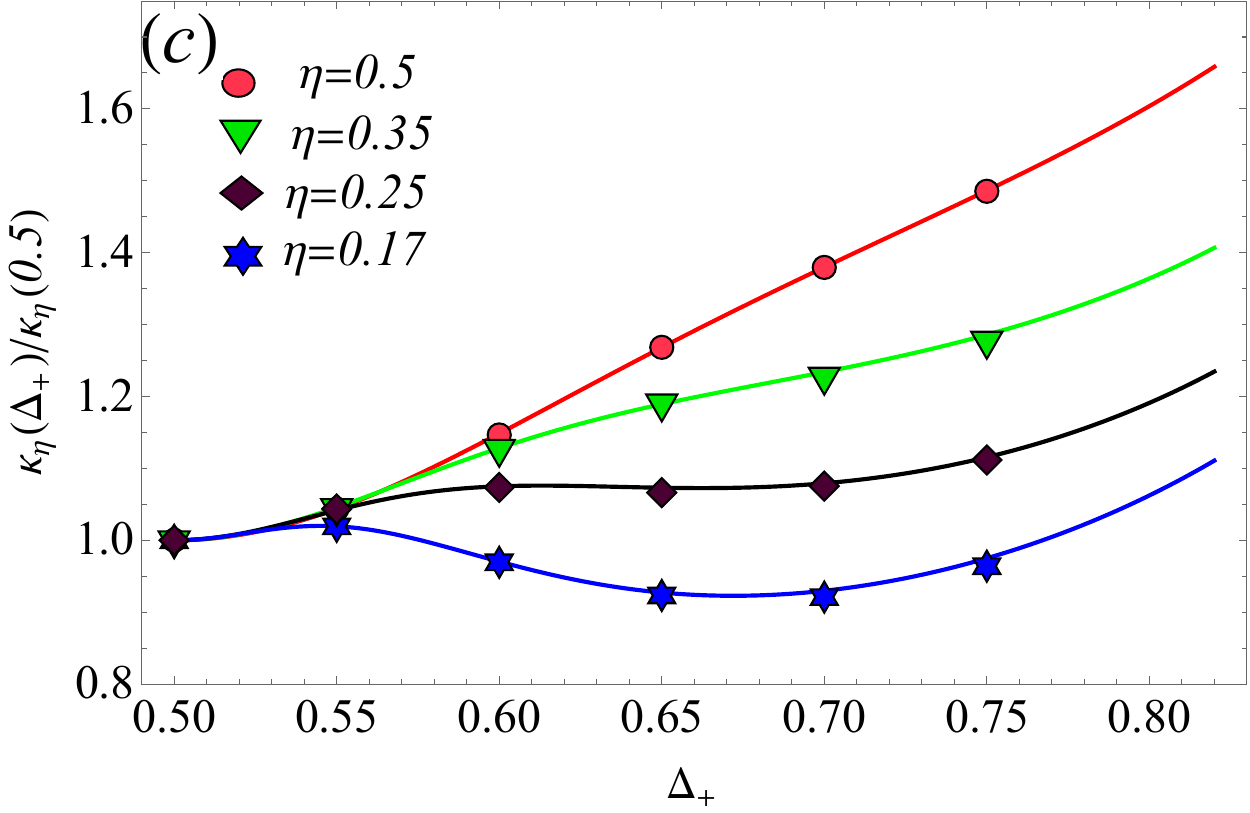}
  \caption{ Various moments for the adder scenario as a function of the noise strength $\langle \eta^2 \rangle$ and the asymmetry $\Delta_+$. Symbols are results of numerical simulations while the thick lines are theoretical results due to Eq.~\eqref{eq:v momenets}.
  {\bf (a)} The first moment $\bigcirc$, and second moment  $\bigtriangledown$, of the cell size $v$ divided by  $e^{\frac{2}{3}\langle \eta^2\rangle N^2}\Gamma[1/2+3\ln(2)/2]$, with $N=1$ for the first moment and $N=2$ for the second. The constant value of the plotted quantity is predicted by Eq.~\eqref{eq:v momenets}. The values $\Delta_+=0.7$ and  $C=1$ were used.
  {\bf(b)} Similar to panel {\bf (a)} but as a function of $\Delta_+$ while $\eta=0.75$.
  {\bf(c)} The kurtosis $\kappa_\eta(\Delta_+)$ (see Eq.~\eqref{eq:kurtosis def}) as a function of $\Delta_+$ normalized by the kurtosis of a log-normal distribution $\kappa_\eta(0.5)$. Notice the appearence of non-monotnic behavior for sufficiently low level of noise strength $\eta$.
  }
  \label{fig:adder moments} 
\end{figure*}

As previously discussed, the adder model characterizes a growth mechanism in which cells consistently augment their size by a fixed quantity during each division event. In the context of the adder model, the affine linear approximation, given by Eq.~\eqref{eq:affine-linear}, corresponds to the case when $\alpha=1/2$. 
The middle panel of Fig.~\ref{fig:timer-example} displays $1000$ time-lines for $a=\ln(v)$ that follow the adder growth law with asymmetric division. From this figure the process seems to converge to a stable behavior where the cells sizes are effectively bounded, unlike the timer model described in Fig.~\ref{fig:timer-example}.

The distribution of cell sizes for asymmetric division is provided by Eq.~\eqref{eq:v stable distribution}.
Figure \ref{fig:adder linear comparisom} presents a nearly perfect agreement between the result of the developed continuous approximation and numerical simulation of the original stochastic map that describes the adder model.


The mathematical form for the moments of the cell sizes, $\langle v^N \rangle = \int_0^\infty v^N {\tilde P}_{asym}(v)\,dv$ 
is easily obtained from Eq.~\eqref{eq:k solution with constants} and is explicitly provided in Appendix B, Eq.~\eqref{eq:v momenets}. 
We notice that the quantity
\begin{equation}
    \label{eq:moment noise independence}
   \frac{ \langle v^N\rangle e^{-\frac{N^2\langle \eta^2 \rangle}{4\alpha(1-\frac{\alpha}{2})}}}
   {{\Gamma\left[\frac{1}{2}+\frac{1-\frac{\alpha}{2}}{\alpha}\gamma\right]}
   }
\end{equation}
is independent of the size of the noise $\langle \eta^2 \rangle$. For adder, $\alpha=1/2$ and the protocol $\gamma = \ln(2)$ we verify this result for the first and second moments in Fig.~\ref{fig:adder moments} {\bf(a)}. The correspondence between the simulation results and theory (Eq.~\eqref{eq:v momenets}) is very good. We further explore the behavior of the moments as a function of the asymmetry parameter $\Delta_+$. In Fig.~\ref{fig:adder moments} {\bf (b)} the expression in Eq.~\eqref{eq:moment noise independence} as obtained from simulations is compared to the theoretical prediction, i.e. Eq.~\eqref{eq:v momenets}. 
The correspondence is very good. 
The monotonic decay of the first moment $\langle v\rangle$ in Fig.~\ref{fig:adder moments} {\bf(b)} tells t us that for the adder, the average cell size can only decrease when asymmetric division is imposed.

The previous section presented the transition between uni-modal to bi-modal behavior of $a$ and $x$. Eq.~\eqref{eq:critical gamma} specifies the critical rate $\gamma_c$ below which a transition between bi-modal and uni-modal can be observed (see Fig.~\ref{fig:phase-line}). For the case of adder $\alpha=1/2$ and therefore $\gamma_c=2/3$. 
Since the protocol $\gamma=\ln(2)=0.693\dots>2/3$ works so well for the affine-linear adder scenario, it also suggests that no transition to the bimodal shape of $P(a)$ is possible. The uni-modal phase is the only type that can be obtained for the distribution of $a$ when the affine-linear adder growth model is explored. 
This doesn't mean that the shape of ${\tilde P}_{asym}(v)$ will be always the same. When the phase-transition line in Fig.~\ref{fig:phase-line} is approached some modifications of ${\tilde P}_{asym}(v)$ emerge. Specifically, in Fig.~\ref{fig:adder linear comparisom} we observe the appearance of a ``shoulder" for ${\tilde P}_{asym}(v)$. Such behavior should be recorded also when exploring the kurtosis of the PDF. The kurtosis 
\begin{equation}
    \label{eq:kurtosis def}
    \kappa_{\eta}\left(\Delta_+\right)=\frac{\langle \left[v-\langle v\rangle\right]^4\rangle}{\langle \left[v-\langle v\rangle\right]^2\rangle^2}
\end{equation}
can be readily obtained by utilizing the moments of $v$, as provided by  Eq.~\eqref{eq:v momenets}.
In Fig.~\ref{fig:adder moments} {\bf (c)} the behavior of the kurtosis is plotted as a function of the asymmetry parameter $\Delta_+$ for various values of the noise $\eta$. For the symmetric division case of $\Delta_+=1/2$, the kurtosis $\kappa_\eta(0.5)$ attains the value of kurtosis of a log-normal distribution, i.e. $\kappa_\eta(0.5)=e^{4\langle \eta^2\rangle/\alpha(1-\alpha/2)}+2e^{3\langle \eta^2\rangle/\alpha(1-\alpha/2)}+3e^{2\langle \eta^2\rangle/\alpha(1-\alpha/2)}-3$.  
For sufficiently small values of $\eta$ the kurtosis shows a non-monotonic behavior, there is even a range of $\Delta_+$s for which the kurtosis is decaying. 
This phenomena can be comprehended by examination of Fig.~\ref{fig:phase-line}.
Since $\gamma=\ln(2)$ the limit of weak noise, i.e. $\eta$, leads to a position in the phase space that is very close to the phase line determined by Eq.~\eqref{eq:phase separation}.
While the bi-modal phase is not reached, the distribution $P(a)$ has a very flat/constant part (see the inset of Fig.~\ref{fig:adder linear comparisom}). 
At this part, ${\tilde P}_{asym}(v)$ decays as $1/v$ and a "shoulder" is developed.  Therefore the observed decay of the kurtosis to small values. 
We also emphasize the strong alignment between the analytic theory of kurtosis and the simulation results, further enhancing the validity of our findings.

 \begin{figure}
\includegraphics[width=0.93\linewidth]{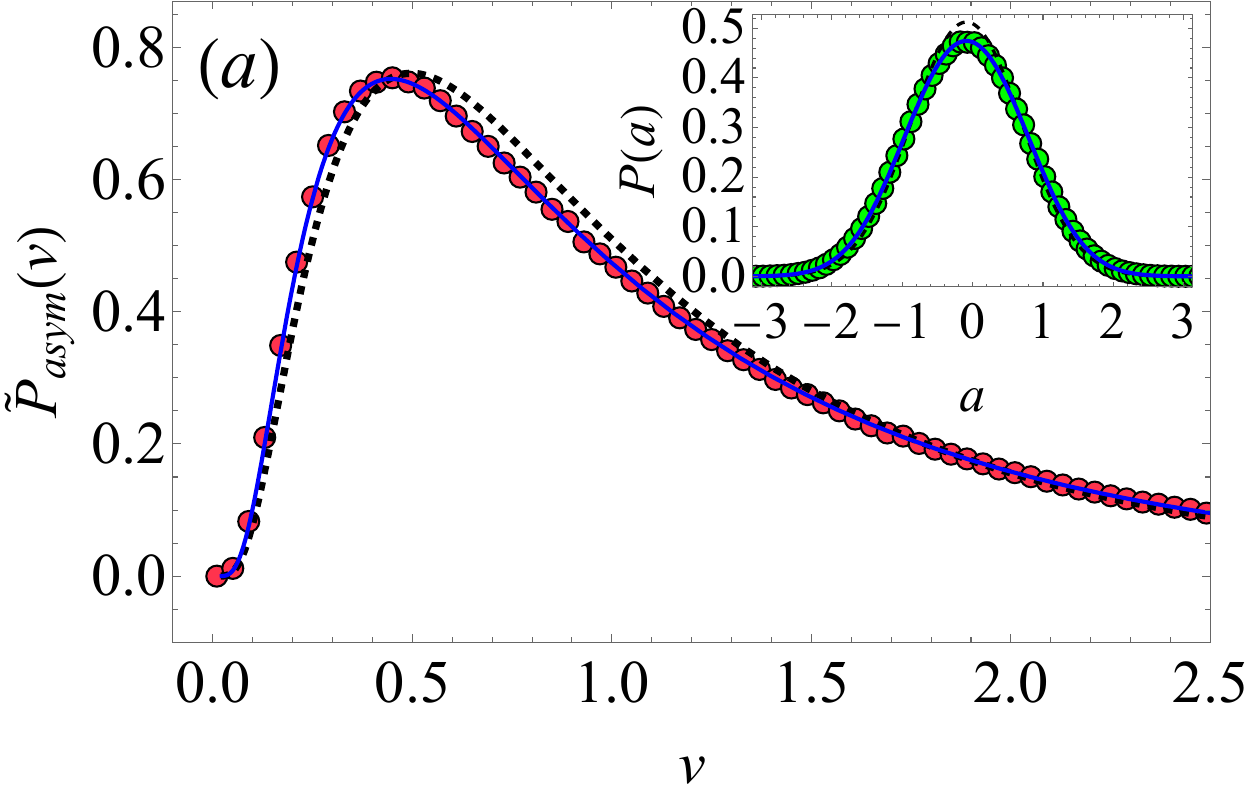}
 \\

    \includegraphics[width=0.93\linewidth]{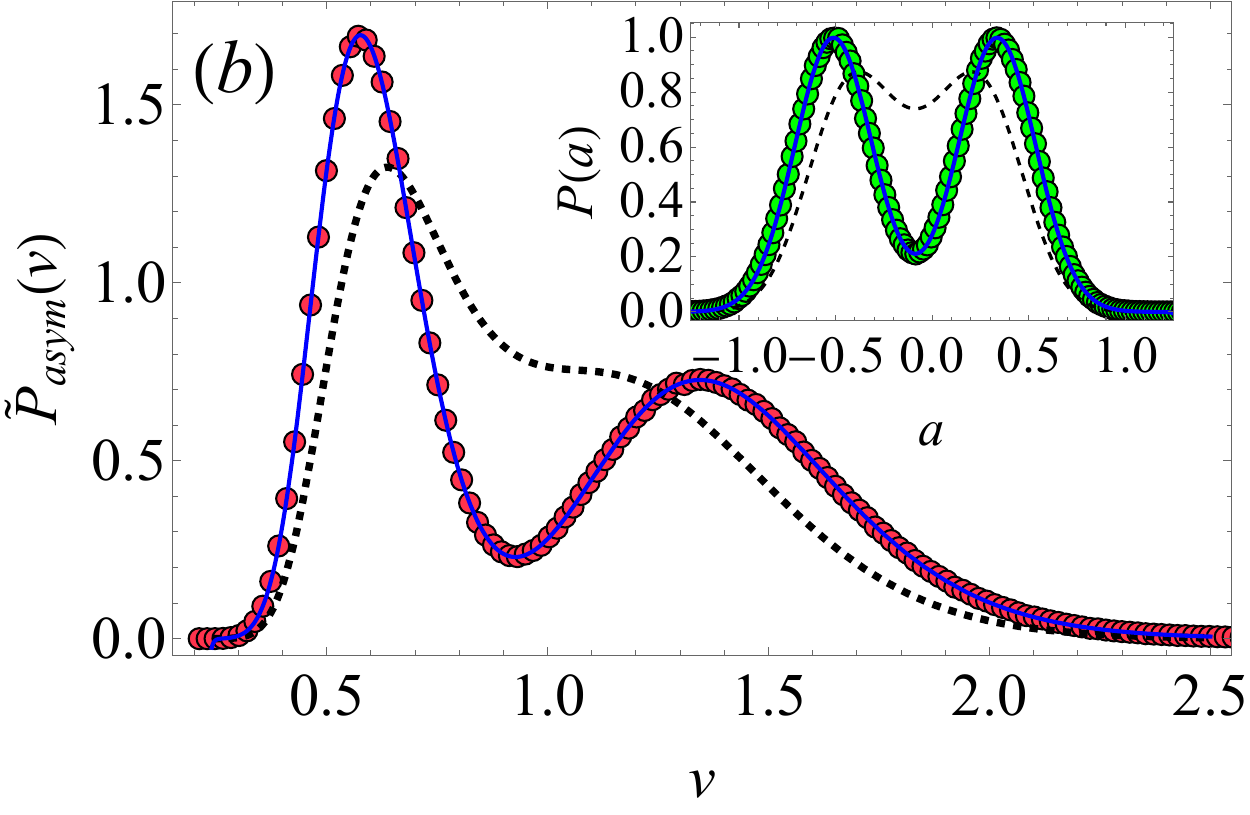}
    \vspace{-4mm}
 \caption{ Stable distributions of the cell sizes $v$ for the sizer scenario while the insets display the distributions of $a=\ln(v)$. The symbols $\bigcirc$ present the results of numerical simulations, dashed lines are the theory, Eq.~\eqref{eq:a stable distribution} and Eq.~\eqref{eq:v stable distribution} with $\gamma=\ln(2)$ and thick lines are for $\gamma=0$.
 For panel {\bf(a)} the parameters $\Delta_+=0.7$, $\eta=0.7$ and $C=1$ were employed, while for panel {\bf(b)} $\eta=0.2$. According to Eq.~\eqref{eq:sizer bimodal transition} this difference in $\eta$ triggers the transition between uni-modal phase in {\bf(a)} to bi-modal phase in {\bf(b)}.}
  \label{fig:sizer PDF} 
\end{figure}

\subsection{Sizer Model}

The sizer model states that division occurs when the cell reaches a critical size.
For the affine-linear approximation the sizer model is the case when $\alpha=1$. When the division is symmetric, Eq.~\eqref{eq:sym solution} describes the PDF of $a$ and cell sizes distribution is log-normal.
The right panel of Fig.~\ref{fig:timer-example} displays the timelines of the sizer model. 
Similar to the adder model, the time-lines of the sizer model seem to converge to a stable distribution.
In Fig.~\ref{fig:sizer PDF} we compare the theoretical results (symbols) to the theoretical prediction due to Eq.~\eqref{eq:v stable distribution} with $\alpha=1$ and $\gamma=\ln(2)$ (dashed line). 
For the larger size of the noise (panel {\bf(a)} the differences are small, while for the smaller size of the noise (panel {\bf(b)}) they are big and non-negligible. 
These differences suggest that the developed approach of continuous approximation of the stochastic map doesn't suit the discrete behavior, at least in the weak noise limit.

While the continuous approximation failed when using the protocol with $\gamma=\ln(2)$, modifying the parameter $\gamma$ to $\gamma=0$ produced a perfect fit of the theory to numerical simulations (thick line in panels {\bf(a)} and {\bf(b)} of Fig.~\ref{fig:sizer PDF}). 
Such modification of $\gamma$ is not simply a lucky guess but rather a consequence of proper inspection of the stochastic map.
For the asymmetric sizer scenario, the behavior of $a_n$ is determined by
\begin{equation}
    \label{eq:sizer stochastic map}
    a_{n+1} = \ln(2C) +\ln(\Delta_\pm)+\eta_n
\end{equation}
meaning the value $a_{n+1}$ is completely independent of $a_n$, and there is no memory in the process. 
The continuous  stochastic approximation employed in this work is provided by Eq.~\eqref{eq:Langevin asymmetric} with $\alpha=1$ and is an Ornstein-Uhlenbeck process with a jumping minimum of the quadratic potential, i.e. dichotomous noise. For such a process of jumping minimum, memory is always present. 
After each jump, the new initial position will be the position of $a_t$ just before the moment of the jump. 
The only way to exclude the memory is to set the jump rate $\gamma$ to $0$. 
In such a way, when the process is observed at long times, it is either at the state with $\Delta_+$ or $\Delta_-$, irrespective of where it was located just a moment ago. 
This approach works perfectly for the sizer as the comparison in Fig.~\ref{fig:sizer PDF} shows (thick lines).

We previously suggested the protocol of $\gamma=\ln(2)$ when modeling the asymmetric division by dichotomous noise for any $0<\alpha<1$. 
Our argument was that the average duration of the number of generations in the $+$ or the $-$ state is the only necessary quantity. 
The sizer example shows that it is generally not true and that the transition rate $\gamma$ of the dichotomous noise should depend on $\alpha$. 
The ``amount" of memory from one generation of bacteria to successive generations affects the rate $\gamma$ of the dichotomous noise.

One last thing to notice is that by switching to the $\gamma=0$ protocol, we effectively created the possibility of transition from uni-modal to bi-modal behavior. 
According to Eq.~\eqref{eq:phase separation} the uni-mdal to bi-modal transition occurs when $_1F_1\left[\frac{3}{2},\frac{1}{2},-\frac{{\tilde\delta}^2}{4{\tilde T}{\tilde\alpha}^2}\right]=0$. When substituting all the parameters and using the fact that $_1F_1[\frac{3}{2},\frac{1}{2},-B^2]=(1-2B^2)e^{-B^2}$~\cite{abramowitz} we obtain the condition for the sizer scenario  to show a bi-modal behavior:
\begin{equation}
    \label{eq:sizer bimodal transition}
    \Delta_+ > \frac{1}{1+ e^{-\frac{1}{2}\langle \eta^2 \rangle}}.
\end{equation}
The presence of the noise strength $\langle \eta^2 \rangle$ in the exponential term accentuates the profound blurring impact of the noise, which obscures the intricate details of asymmetric division that lie beneath.



\section{Summary}
\label{chapter: Discussion}
How cells regulate their size and achieve size homeostasis is a subject of interest and importance in Biology. 
Recent single-cell studies led to extensive contributions and advancements in the topic. 
Several mathematical approaches were constructed to describe the growth and division of symmetrically dividing organisms.
Of specific interest and applicability is the approach that utilizes the stochastic map description of the growth and division process\cite{amir}. 
The adder model was successful in describing experimentally observed cell size distribution of different bacteria such as {\it E. Coli}~\cite{Suckjoon2016} and budding yeast~\cite{Soifer2016}.  

The main emphasis of this study is centered on extending the utilization of the stochastic map approach to account for asymmetric division. While previous studies~\cite{marantan&amir} have explored this topic, a closed-form solution has yet to be established, to the best of our knowledge. To tackle this challenge, we utilize the affine-linear approximation for the growth laws scenarios and expand upon the existing formalism of a second-order continuous approximation~\cite{kessler&burov, burovArxiv} of the discrete stochastic map.
One key advantage of employing the affine-linear approximation in modeling the growth law is the incorporation of the parameter $\alpha$.
Through modifications to $\alpha$, the growth law exhibits a transformative behavior, enabling a shift across various scenarios, including timer, adder, and sizer.

In our research, we establish a comprehensive connection between the phenomenon of asymmetric division and the Ornstein-Uhlenbeck process with dichotomous noise, i.e.,  a stochastically jumping minimum of the quadratic potential.
Remarkably, we succeed in obtaining an exact mathematical expression for the PDF associated with this process.
By leveraging this mathematical framework, we demonstrate the existence of two distinct characteristic behaviors, namely bi-modal and uni-modal distributions.
Additionally, we were able to derive an explicit mathematical representation of the phase separation line, providing further insights into the underlying dynamics.
To validate the efficacy of our approach, we extensively compare our analytical results to numerical simulations encompassing timer, adder, and sizer scenarios. 
In the timer scenario, similar to symmetric division, we find that no stable distribution exists.
Our analytical approximation exhibits an excellent fit to the adder scenario, showcasing the robustness of the developed method in capturing its dynamics accurately.
For the sizer scenario, we encountered the need to introduce a modification to the rate of the dichotomous noise to achieve a satisfactory correspondence between our analytical results and the observed behavior.

The existence of a bi-modal state for the cell division process is visible for the sizer model where the condition for the bi-modal phase is provided by a very simple formula, i.e., Eq.~\eqref{eq:sizer bimodal transition}. 
In the case of the adder, we demonstrate that a transition to a bi-modal state is not possible, and only the uni-modal phase prevails. 
Despite the nonmonotonic behavior of the kurtosis (Figure \ref{fig:adder moments} (c)), which indicates the presence of a ``shoulder" in the distribution of $v$, our simulations did not reveal any bi-modal shape for $v$ or $a$, aligning with our theoretical predictions.
This finding appears to contradict the observations made by Marantan and Amir~\cite{marantan&amir}, who documented the presence of a bi-modal shape in the adder model. The discrepancy in results stems from the distinct forms of the growth law employed. 
While our study utilized the affine-linear approximation, the previous investigation employed the non-linear adder scenario. 
This disparity, coupled with numerous other observed distinctions~\cite{Grilli2017}, is noteworthy. The affine-linear approximation corresponds to an Ornstein-Uhlenbeck process, whereas the non-linear adder is associated with a process featuring an asymmetric and non-quadratic effective potential~\cite{burovArxiv}.

While the impressive ability to accurately represent the cell size distribution in closed mathematical form is noteworthy, the true strength of the presented approach lies in its mapping to an Ornstein-Uhlenbeck process with dichotomous noise. 
By expressing the biological parameters in terms of temperature, force, and potential, as we have done, possible intriguing insights about the system can be deduced~\cite{Mugler2023}. For example, the advantages or disadvantages of asymmetric division can be examined in the context of the stability of a physical process and its resilience to rare events, such as outliers.
Another direction is the effect of asymmetry in division on population growth~\cite{barber2021}.
Furthermore, the presented approach has the potential to be extended to encompass organisms that undergo multiple fission, such as certain protists (e.g., sporozoans and algae). 
This extension will involve replacing dichotomous noise with other types of noises, like trichotomous noise \cite{mankin} or other telegraphic jump processes \cite{othmer}. Additional cases where the further extension of the developed mapping can be applicable are combinations of growth laws ~\cite{banerjee2017, Graziano2018} and presence of correlations~\cite{Brenner2018}.

{\bf Acknowledgments: }This work was supported by the  Israel Science Foundation Grant No. 2796/20.

\appendix
\section{Appendix A}
\setcounter{equation}{0}
\renewcommand{\theequation}{A.\arabic{equation}}

In this section, we describe how the decoupling of the coupled Fokker-Planck equations described by Eq.~\eqref{eq:coupled fokkerP} is performed~\cite{dybiec}.
First we take the $t\to\infty$ limit and assume that the process has reached a steady state, namely $\frac{\partial P_{\pm}(x,t)}{\partial t} = 0.$ The outcome is a pair of coupled equations
\begin{equation}
    \label{eq:pplus coupled}
    T P_+''(x)+\frac{\partial}{\partial x} 
    \left[\left(U'(x)+{\tilde \delta}\right)P_+(x)
    \right]
    -\gamma P_+(x)+\gamma P_-(x)=0
\end{equation}
\begin{equation}
    \label{eq:pminus coupled}
    T P_-''(x)+\frac{\partial}{\partial x} 
    \left[\left(U'(x)-{\tilde \delta}\right)P_-(x)
    \right]
    -\gamma P_-(x)+\gamma P_+(x)=0.
\end{equation}
We define $P(x)=(P_+(x)+P_-(x))/2$ as the PDF of obtaining the value $x$ and its complimentary $Q(x)=(P_+(x)-P_-(x))/2$. 
Addition and subtraction of Eq.~\eqref{eq:pplus coupled} and Eq.~\eqref{eq:pminus coupled} results in
\begin{equation}
    \label{eq:pandq first}
    \frac{\partial }{\partial x}
    \left[
\left(
T\frac{\partial}{\partial x}
+U'(x)\right)P(x)+{\tilde \delta}Q(x)
    \right]
    =0
\end{equation}
\begin{equation}
    \label{eq:pandq second}
    \frac{\partial }{\partial x}
    \left[
\left(
T\frac{\partial}{\partial x}
+U'(x)\right)Q(x)+{\tilde \delta}P(x)
    \right]-2\gamma Q(x).
    =0
\end{equation}
Integration of Eq.~\eqref{eq:pandq first} leads to 
\begin{equation}
    \label{eq:q(x) solution}
    Q(x)=-\frac{1}{\tilde \delta}
    \left[ 
    T\frac{\partial P(x)}{\partial x}+U'(x)
    \right],
\end{equation}
where we utilized the condition $P_{\pm}(x)\to0$ when $|x|\to\infty$.
Substitution of Eq.~\eqref{eq:q(x) solution} into Eq.~\eqref{eq:pandq second} leads to Eq.~\eqref{eq:3rd order ode}.

\appendix
\section{Appendix B}
\setcounter{equation}{0}
\renewcommand{\theequation}{B.\arabic{equation}}

In this Appendix, the analytic expression for the moments of cell sizes $\langle v^N\rangle$ is developed. The affine-linear approximation for growth laws, i.e. Eq.~\eqref{eq:affine-linear}, and asymmetric division are assumed. 
We use the fact that 
\begin{equation}
\langle v^N \rangle = \langle e^{Na} \rangle = \langle e^{N\left(x + \ln(C) + \ln(2)/\alpha + [\ln(\Delta_+) + \ln(\Delta_-)]/2\alpha\right)} \rangle
    \label{eq:mommnets connection}
\end{equation}
and
\begin{equation}
\langle e^{-ikx}\rangle = {\hat P}(k)
    \label{eq:furie connection}
\end{equation}
to obtain from Eq.~\eqref{eq:k solution with constants}
\begin{equation}
\begin{array}{l}
\langle v^N \rangle =\left(C \left(2\sqrt{\Delta_+(1-\Delta_+)}\right)^{\frac{1}{\alpha}}\right)^N
\Gamma[\frac{1}{2} + \frac{(1-\frac{\alpha}{2})}{\alpha}\gamma] 
\times
\\
\left({\ln\left(\frac{\Delta_+}{1-\Delta_+}\right)}\frac{ N}{4 \alpha}\right)^{\frac{1}{2}-\frac{(1-\frac{\alpha}{2})}{\alpha}\gamma} e^{\frac{N^2 \langle \eta^2 \rangle}{4 \alpha(1-\frac{\alpha}{2})}} 
\times
\\
I_{(\frac{1}{\alpha}-\frac{1}{2})\gamma-\frac{1}{2}}\left( {\ln\left(\frac{\Delta_+}{1-\Delta_+}\right)}\frac{N}{2 \alpha}\right).
\end{array}
    \label{eq:v momenets}
\end{equation}
where $I_\omega()$ is the modified Bessel function of the first kind~\cite{abramowitz}.

\bibliography{./bacteriaref.bib}

\end{document}